\documentclass[pra,aps,floatfix,showpacs,superscriptaddress]{revtex4-1}
\usepackage{epsfig,graphicx,tabularx}
\pdfoutput=1
\begin{document}
\title{ Dark soliton collisions  in a toroidal Bose-Einstein condensate}
\author{D. M. Jezek}
\affiliation{IFIBA-CONICET}
\author{P. Capuzzi}
\affiliation{IFIBA-CONICET}
\affiliation{Departamento de F\'{i}sica, Facultad de Ciencias Exactas y Naturales, Universidad de Buenos Aires,
Pabell\'on 1, Ciudad Universitaria, 1428 Buenos Aires, Argentina}
\author{H. M. Cataldo}
\affiliation{IFIBA-CONICET}

%
\begin{abstract}

We study the dynamics of two gray solitons in a Bose-Einstein condensate confined by a toroidal trap
with a tight confinement in the radial direction. Gross-Pitaevskii simulations show
that solitons
can be long living objects passing through many collisional processes. We have observed quite
different behaviors depending on the soliton velocity.
Very slow solitons, obtained by perturbing the stationary solitonic profile, 
move with a constant angular velocity until they collide elastically and move
in the opposite direction without showing any sign of lowering their energy. 
In this case the density 
notches are always well separated and the fronts are sharp and straight. 
Faster solitons present 
vortices around the notches, which play a central role during  the collisions. 
We have found that in these processes the solitons lose energy, as the outgoing
velocity turns out to be larger than the incoming one.
To  study the dynamics, we model the gray soliton state with
a free parameter that is related to  the soliton velocity. We further analyze the energy,  soliton velocity and 
turning points in terms of such a free parameter, finding that the main features are in accordance with
the infinite one-dimensional system.

\end{abstract}

\maketitle
\section{Introduction}

Topological defects has been a central topic in nonlinear systems of various fields in physics.
In Bose-Einstein condensates (BECs), such defects include vortices, solitons and solitonic vortices (svortices).
Solitons are characterized by their form stability under time evolution and can behave akin to classical 
particles.
For a repulsive interaction between atoms, black solitons are stationary states with a $ \pi $ jump in the 
phase
of the order parameter, which produces a density nodal surface. 
 In contrast, gray solitons are moving objects with a nonvanishing density dip, characteristic of 
a smaller phase difference between both sides of the density notch. 
In an infinite one-dimensional (1D) system the 
collision of solitons has been thoroughly studied from the theoretical viewpoint \cite{tsu71,zak73,libro1,theo10}.
  It has been shown    
that for soliton speeds smaller (larger)
than  half  the sound velocity, the solitons remain separated (overlapped) at the collision,
appearing  to be {\it reflected} by ({\it transmitted} through)  each other
\cite{theo10}.
 In such a system the solitons collide elastically and continue moving with a constant velocity away from  the collision region.
In particular, for very slow solitons the system can be safely  regarded  as hard-sphere-like 
particles that interact through an effective (velocity dependent) repulsive potential \cite{theo10}.

On the other hand, vortices are characterized by a quantized circulation of the velocity field around the position 
where the density vanishes.
Svortices  \cite{brand02} are present in tightly confined systems, and differ from standard  vortices in the 
form of 
their density distribution which looks 
quite similar to 
the soliton one, although the velocity field  changes its sign along the density dip,  where the vortex is located.
 
Solitons in  atomic BECs confined with different trapping potential geometries 
 have been extensively studied   in the last years 
\cite{ libro,rev10}. And renewed interest has arisen from the experimental observation of solitonic vortices in
bosonic and  fermionic systems \cite{donadello14,ku14,chevy14}. In such recent  BEC 
experiments, solitons have been spontaneously created  through the Kibble-Zurek mechanism \cite{lamporesi13}.

The  commonly used candidate  to experimentally study the soliton dynamics in a quasi 1D system
 has been a cigar-shaped condensate \cite{burger}. 
However, due  to the  harmonic trapping potential,  such a single soliton dynamics
differs considerably with respect to that of the strictly 1D case, where the soliton moves with
a constant speed. That is, in the Thomas-Fermi approximation a soliton oscillates in a cigar-shaped condensate
with a frequency $ \omega_s= \omega_{trap} / \sqrt{2}$ \cite{fedi99,anglin,konotop04,theo10}, 
where $\omega_{trap}$ is the angular 
frequency
of the trap in the longitudinal direction,
a result that can be interpreted in terms of the definition of the soliton mass
\cite{becker08}. It is worthwhile noticing 
that one can avoid  such a potentially undesirable effect stemming from the harmonic
trap by utilizing a toroidal-shaped
condensate.
 However, only few works  have undertaken the study of
  soliton dynamics in toroidal condensates including the possible formation of svortices 
\cite{brand01,muntsa15}.

The aim of this work is to study the double-notch soliton dynamics occurring in a toroidal BEC, 
which involves many collisional processes with a related vortex dynamics.
In Section II we introduce the system, particularly the toroidal trap and the set of parameters involved.
In Section III, first we numerically obtain by solving the stationary Gross-Pitaevskii (GP) equation the black
 soliton order parameter. 
In a second step, based on the 1D black soliton profile
 we construct gray solitons with imprinted velocities
which range from very slow values up to velocities near the ground-state sound speed.
By solving the time-dependent 
GP equation,
we study in Section IV  the dynamics of such gray solitons, 
observing that there exist two different regimes depending on the
type of collision involved and the role  played by vortices.
  Section V is devoted to the analysis of the energy, soliton velocity and turning points, 
where we discuss their behavior in comparison to that of the infinite 1D system.  Finally, the 
conclusions of our study are gathered in Section VI.

\section{The system}

The trapping potential is written as the sum of a term $V_{\text{RG}}$ that 
depends on the radial coordinate $r=\sqrt{x^2+y^2}$ and gives rise to the
toroidal shape of the condensate,
and a term that is harmonic in the $z$ direction:
\begin{equation}
V_{\text{trap}}(x,y,z)=V_{\text{RG}}(r) + \frac{1}{2} M  \omega_z^2  z^2 \, ,
\end{equation}
  where  $M$ denotes the atomic mass of $^{87}$Rb and $\omega_z$ is
the trap frequency in the $z$-direction.
The  potential term that confines the atoms in  the radial direction  is modeled as the following  ring-Gaussian potential 
\cite{mu13}
\begin{equation}
V_{\text{RG}}(r)=    -V_0 \, \exp\left[ - \Lambda \left(\frac{r}{r_0}-1 \right)^2\right],
\label{toro}
\end{equation}
where $V_0$ and  $r_0$   denote the depth and  radius 
of its  minimum. 
The dimensionless parameter $\Lambda $ is associated to the $1/e^2$ width ( $ w = r_0 \sqrt{ \frac{ 2}{\Lambda} }$)  of such a
 ring-Gaussian potential.

The trap parameters have been selected according to the 
experimental conditions of
 Ref. \cite{boshier13}. We have set  $V_0$=70 nK, $r_0$ = 4 $\mu \text{m} $, and we will work with a fixed particle number 
$ N=3000$.
We will further assume a high value of 
$\omega_z=2\pi\times 921.77$ Hz, yielding
a quasi two-dimensional condensate that allows a simplified numerical treatment \cite{castin}.
Then,  the order parameter can be represented as a product
of a wave function on the $x$-$y$ plane $\psi(x,y)$,
and a Gaussian wave function along the $z$ coordinate  from which
the following effective two-dimensional interacting  parameter can be extracted  \cite{castin}
\begin{equation}
g=g_{3D}\left(\frac{M\omega_z}{2\pi\hbar}\right)^{1/2},
\end{equation}
where $g_{3D}=4\pi\hbar^2a/M$,
$a= 98.98\, a_0 $ being the  
$s$-wave scattering length of $^{87}$Rb and $a_0 $ the Bohr radius.
The value of the remaining dimensionless parameter $\Lambda=20$ ($ w = 1.265 $ $\mu \text{m}$) was chosen to assure also a 
tightly confined condensate in the radial direction.
 However, we have not simplified our treatment to 
a 1D system because, as we will see,  vortices play an interesting role in the dynamics.
Finally, it is worth mentioning that henceforth all the 
 order parameters will be  normalized to the number of particles.

\section{Dark  solitons }

\subsection{Black soliton}

\begin{figure}
\includegraphics{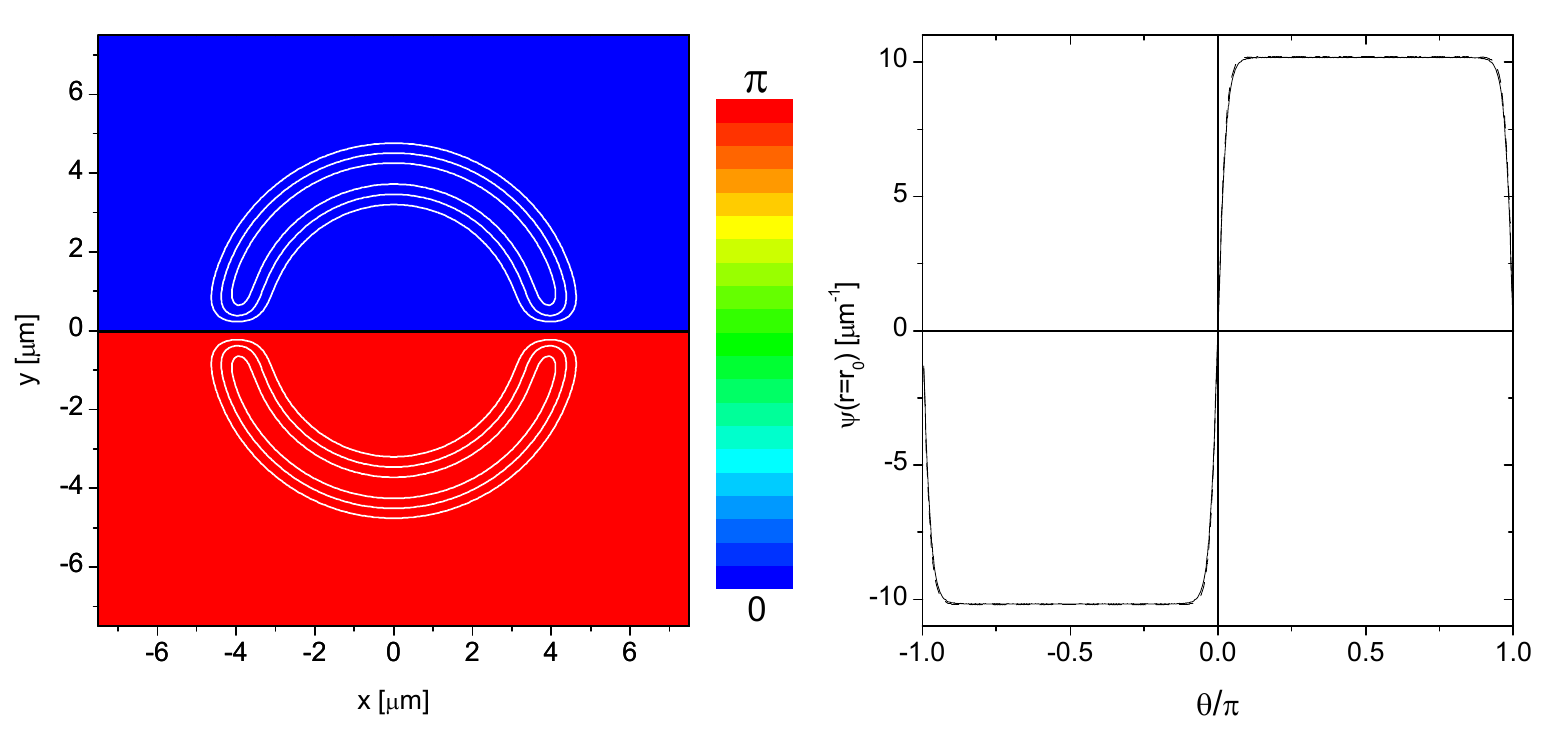}
\caption{(Color online) Black soliton isodensity contour (solid white lines) and phase 
distribution (colors) are shown in the left panel. 
The GP order parameter (solid line) and the almost superposed results from
 Eq. (\ref{wfps}) with $f(r= r_0)= 1$ and $k=2.785$ $\mu$m$^{-1}$
(dotted line)  are displayed on 
the right panel.}
\label{figu1n}
\end{figure}

To obtain the exact GP  black soliton order parameter,  we numerically seek for 
a state which exhibits  uniform phases at both sides of a nodal straight line
with a $\pi$ jump between them. 
We note that, expressed as a function  of   the angle  $ \theta $ defined  along the torus, such an  order parameter presents 
 two density notches located  at $ \theta =0 $ and $ \theta =\pi $, and thus can be treated as a two-soliton system.
The structure of such a state is in accordance with the 1D picture, in which dark solitons interact with each other 
via a repulsive potential \cite{theo10}. In fact, one expects that in a stationary configuration both solitons 
should be separated from each other by the largest
distance compatible with the constraints, 
which in this case corresponds to diametrically opposite positions along the ring.
We have numerically calculated  the GP order parameter of this double-notch  soliton by applying
 an imaginary time method to an initial state given by the  ground state order parameter 
with an imprinted  phase 
difference of $ \pi $ between the half-planes delimited
  by the $x$ axis. 
 The resulting  solution of the  stationary  GP equation   presents a 
nodal line  along the $x$ axis,  preserving the phase difference of $ \pi $.
In Fig.~\ref{figu1n} we show the corresponding phase distribution and  isodensity contour (left panel),
along with the real order parameter  as a function of 
 $ \theta  $ at $ r= r_0$ (solid line, right panel).

Due to the tight confinement in the radial direction, one expects that the double-notch
black soliton should present a similar profile in the angular direction
to that of the two-node stationary analytic solution in a strictly 1D ring system,
which is given in terms of Jacobi elliptic functions \cite{carr}.
As stated in such reference, since the zeros of the solution are well separated
the analytic behavior near such points approaches a hyperbolic tangent function.  
This is in fact so for our two-dimensional order parameter, since it may be  safely  modeled
by
\begin{equation}
\psi_{B}(r, \theta )= \sqrt{n} \,  f(r)  \,   \text{tanh}( k   r  \sin\theta  ) ,
\label{wfps}
\end{equation}
where $n$ denotes the maximum  value of the particle density,  $k$ a parameter of the order of the inverse of
the healing length, and $f(r)$ represents a function to
be determined. 
Particularly, in the right panel of Fig.~\ref{figu1n}, we may appreciate the agreement with the GP result
for $f(r= r_0)= 1$ and the value $k=2.785$ $\mu$m$^{-1}$ which will be  obtained
in the following subsection.

\subsection{Gray soliton}

To  study the solitonic dynamics we may construct a gray soliton
 order parameter
by introducing an imaginary part in the black soliton state (\ref{wfps}). More precisely,
in analogy with the 1D case \cite{tsu71}, we propose  dynamical 
states of the form,  
\begin{equation}
\psi_{G}(r, \theta )= \sqrt{n} \,  \, f(r)  [ \sqrt{1- 
X^2}  \, \text{tanh}(\sqrt{1- X^2}\, k   r \sin\theta) + i X ] \,.
\label{wfp}
\end{equation}
We note  that for a very large torus radius, one can  associate the above state  to  infinite 1D  solitons moving in the
$y= r \sin\theta$  direction with velocity $v_s=cX$, $c$ being the sound speed \cite{tsu71}.

Due to the lack of an analytical expression for $f(r)$, we shall adopt three different proposals
 in order  to  minimize possible residual excitations that could arise at different ranges of the soliton depth,
however,  we will show that the corresponding results do not differ significantly from each other.
 The first proposal consists in dividing the GP black soliton order parameter by  
 $ \text{tanh}( k   r  \sin\theta  )$, next
multiplying the result obtained by  $ \sqrt{1- 
X^2} \,  \text{tanh}(\sqrt{1- X^2} k   r \sin\theta) + i X  $, and finally renormalizing 
to $N$, which we shall call a perturbed black soliton (PBS) state.
The second state, which we will call perturbed ground (PG) state, comes from repeating the above procedure with the ground-state wavefunction 
instead of the black soliton one and omitting the division step.
We expect these approximate order parameters should work satisfactorily not far from $X=0$ ($X=1$)
for the PBS (PG) state.

\begin{figure}
\includegraphics{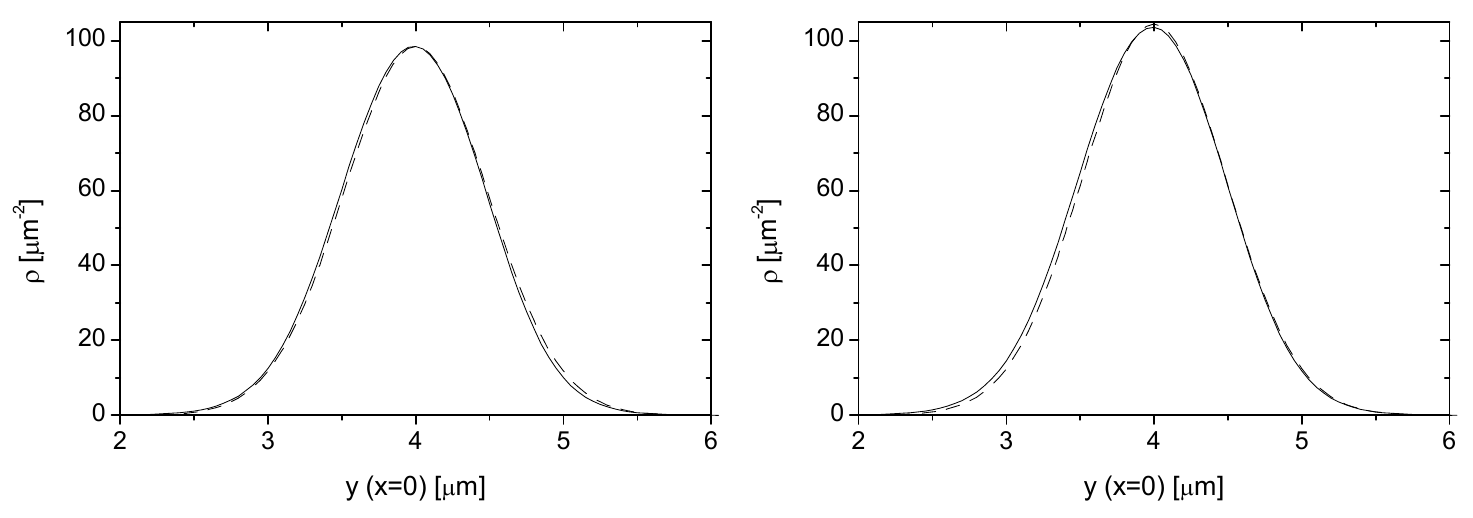}
\caption{Density profiles  $\rho $ as functions of the radial coordinate for the ground (left panel) 
and black soliton (right panel) states. Solid lines correspond to the GP density, whereas
dashed lines correspond to  $ \rho_{GM} = | \psi_{GM}|^2  $, where $\psi_{GM}$ is   given by Eq. (\ref{gaus}) with $\gamma=34$,
for $X=1$ (left panel) and  $X=0$ (right panel). }
\label{figu2n}
\end{figure}

By inspection of  the radial dependence of the order parameters of both ground  and black soliton states, we have
found that they can be approximately modeled with a Gaussian profile.
Then, to cover  a wider range of  $X$ values,  we propose  a third choice that reproduces 
 such a radial dependence. We thus assume  the following order parameter,
\begin{equation}
\psi_{GM}(r,\theta)= \sqrt{n(X)} \, \exp \left[ - \frac{\gamma}{2}\left(1-\frac{r}{r_0}\right)^2\right]
 \,  \left[ \sqrt{1- 
X^2}  \tanh\left(\sqrt{1- X^2}\, k   r \sin\theta\right) + i X \right]  ,
\label{gaus}
\end{equation}
which,   choosing an adequate value of $\gamma$,
  fits quite well the radial profile of the GP density, both for the black soliton ($X=0$)
and the ground state ($X=1$), as can be seen in  Fig.~\ref{figu2n}.  
 We will call this state as the Gaussian model (GM) state.

We note that we have included the  function $n(X)$ in the GM order parameter 
in view that  in a finite system, in contrast to an infinite one,
 the existence of a 
 hole in the condensate density increases the  density maximum with respect 
to that of the ground state. 
We may obtain an analytical  expression  for $n(X)$ by incorporating the constraint that the number of particles for each 
 $X$ value should be equal to that 
of the ground state.
In doing so, due to the tight confinement in the radial direction and  the fact that the width of the notch is much smaller 
than the diameter of the torus,
 in the integral of the density $ \rho_{GM}= |\psi_{GM}(r,\theta)|^2 $   we may safely  replace 
$ r \, \text{sin}\theta$ by $r_0 \,  \theta$ inside the argument of the  hyperbolic function. 
We have verified that such an approximation  introduces an error in  the integral 
of less than one percent. 
Then, under such assumptions  we obtain the following condition, 
\begin{equation}
n(X)= \frac{ n_0}{ 1 - \frac{2 \sqrt{1- X^2}}{ \pi k r_0}},
\label{densidad}
\end{equation}
where $n_0$ denotes the ground-state density  maximum located at $ r= r_0 $, which is fitted to the GP value.
 
The remaining parameter $k$ can be estimated  by applying   the stationary GP equation  to   the black  soliton state (\ref{gaus}),
 which evaluated at  $r=r_0$
yields 
\begin{equation}
k = \frac{\sqrt{n(X=0) g M}}{ \hbar},
\label{densida}
\end{equation}
which turns out to be the inverse of the local healing length at $r=r_0$.
Therefore, combining Eq.  (\ref{densidad}) for $X=0$   and  Eq. (\ref{densida}),
we obtain the value $k=2.785$ $\mu$m$^{-1}$.

Finally, we note that
 the usage of  $n(X)$ in the  GM order parameter for calculating  the density, turns out to be important to 
 accurately  describe 
the density maximum for every $X$ value
in comparison to GP simulations, as can be observed for $X=1$ at the right panel of Fig.~\ref{figu2n}.

\section{The dynamics}

To study the soliton dynamics, we have solved the time-dependent GP equation
 using as an
initial order parameter any of the three alternatives described in the previous section.
By varying the  parameter $X$, which is associated to the initially imprinted soliton velocity, we have observed 
different  dynamics that show a transition around $X=0.5$.

\subsection{Soliton reflection at the collision ($ 0 < X < 0.5 $)}

\begin{figure}
\includegraphics{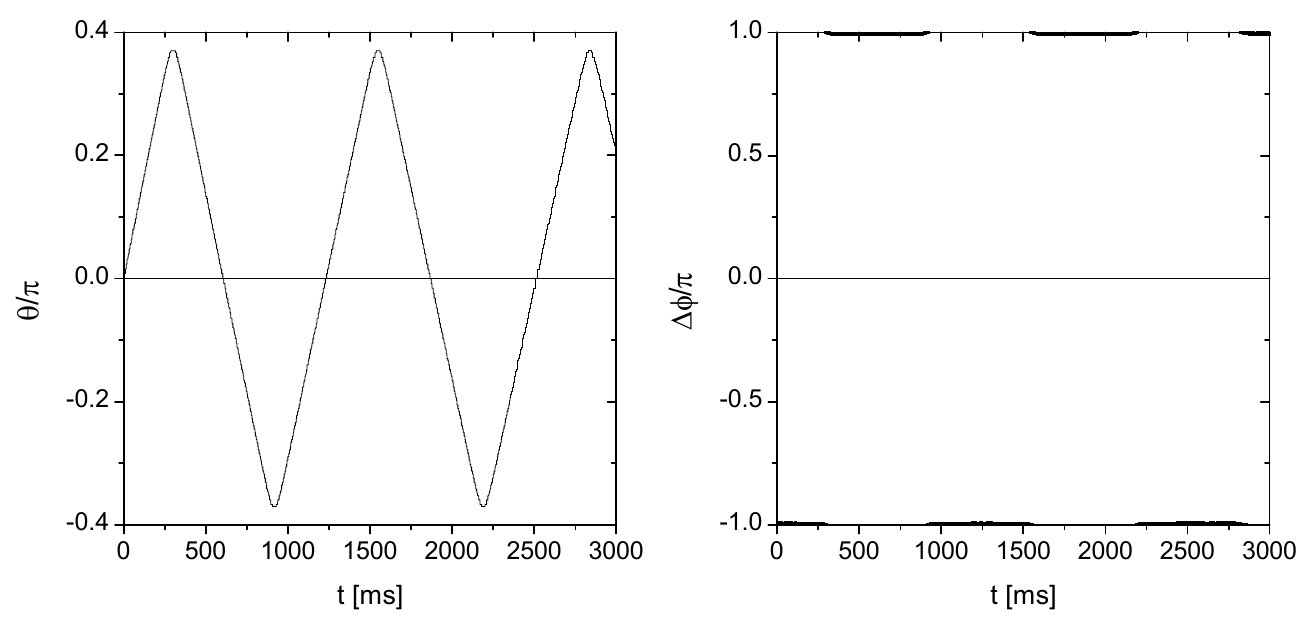}
\caption{Angular position (left panel) of the soliton located at $x > 0$
and the phase difference $\Delta \phi $ (right panel), are depicted
as functions of the time  using an initial PBS profile with the black soliton slightly perturbed with $X=0.01$.}
\label{figub}
\end{figure}

For small values of  $X$ corresponding to low velocities of the defect,  solitons remain  
as   long living objects, 
 whose dynamics presents almost 
constant velocity, except for  very narrow time intervals when the
collisions  occur. In this case the collisions seem to be   elastic.
 On the other hand, the presence of vortices is only observed far away from the condensate, 
and they do not play any role in the 
soliton dynamics. In general, our findings are quite similar to those of the homogeneous 1D system.
  In Fig.~\ref{figub}  we show the results of GP simulations for a PBS initial order parameter 
 with a small  imprinted velocity obtained with $X=0.01$.  
 The angular position  $\theta(t)$  of the 
soliton located at positive $x$ values is depicted at the left panel. 
It may be seen that  the absolute value of the slope of $\theta (t)$
is almost the same throughout the evolution,  except for small intervals around the turning points.
Hence the soliton energy, which depends on its speed, seems to be conserved. 
Another signature of such a conservation comes from the fixed positions of
the turning points that remain located at  $ \theta = \pm 0.37008\, \pi $.
On the other hand, in the right panel we depict
the phase  difference 
between the  upper and lower  regions separated by the notches. 
Specifically, we have evaluated such a phase difference
  as $\Delta \phi = \phi(x=0,y=4\,\mu$m$)-\phi(x=0, y=-4\,\mu$m$)$.
  One can observe that $\Delta \phi(t)$
 alternates between values near $  -\pi$ and $ \pi$ in the intervals of  increasing
and decreasing angles, respectively.

 With such an initial state, which is very
close to the stationary black soliton, one finds that  the system evolves with quite pure solitonic fronts.
\begin{figure}
\includegraphics{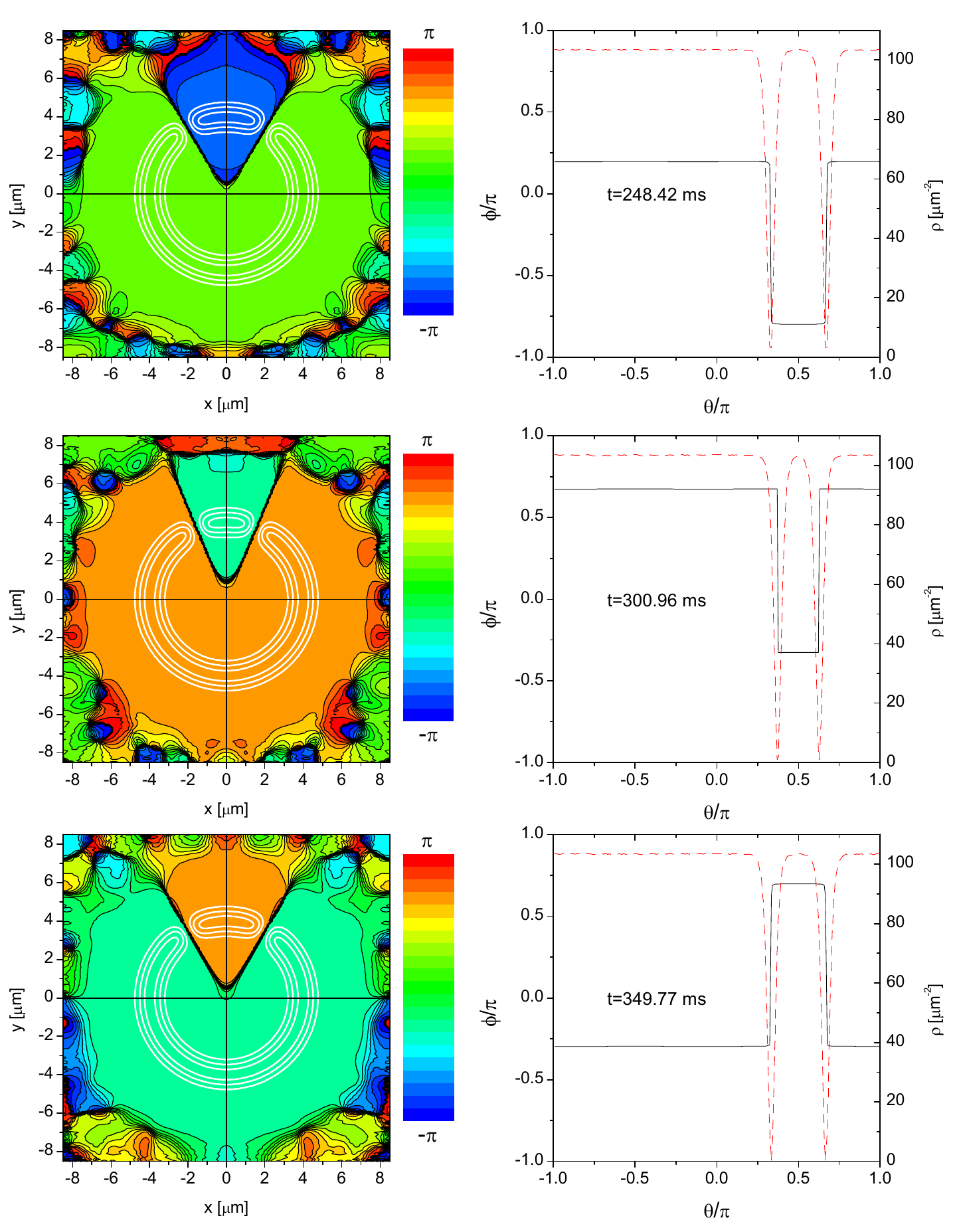}
\caption{(Color online) Snapshots before (top panels), during (middle panels), and after (bottom panels)  the first soliton
collision for the same initial state of 
Fig.~\ref{figub}. In the left panels we show the phase distribution (colors) and density isocontours 
(white solid lines), while in the right panels we depict the
corresponding angular distribution of particle density  (red dashed line)  and phase 
(black solid line) at $r= 4 \, \mu$m.  }
\label{figua}
\end{figure}
In Fig.~\ref{figua}  we show three snapshots of phase and particle density around the first collision (or turning point).
In the top-right panel,
one observes that before the collision the phase gradient is almost zero, except in small regions around the dips,
 where it points
in the opposite direction to the soliton velocity. And,
as expected, only few particles situated at the borders of the notch
 move in this opposite direction.
In the middle-right panel,  it may be seen that during the collision  the density maintains 
two well separated notches, whose minima go to zero, which is
characteristic of a vanishing soliton velocity corresponding to a real order parameter with a phase difference of $\pi$.
At such a turning point, the velocity field is zero even around the notch, as seen from the vanishing phase gradient everywhere, 
except for the  phase discontinuity where the density vanishes. Finally, after the
collision the velocity field is reversed and the soliton continues moving in the opposite direction, as shown at the
bottom panels.

\begin{figure}
\includegraphics{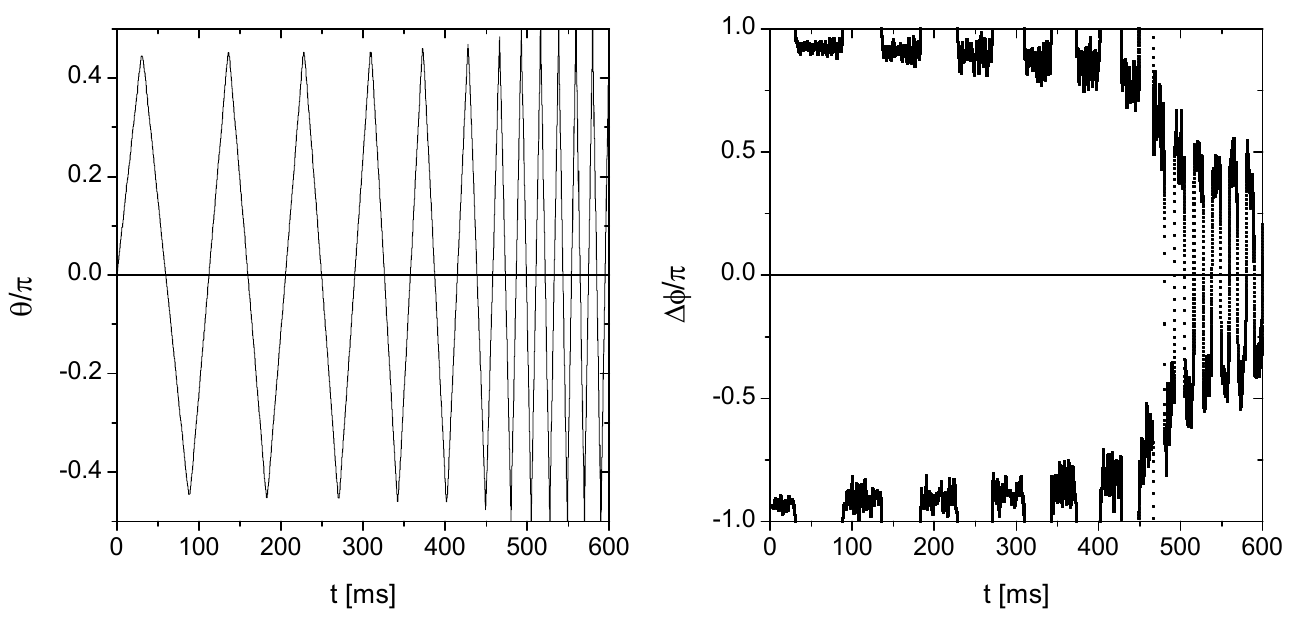}
\caption{Angular position (left panel) of the soliton located at $x > 0$
and the phase difference $\Delta \phi $ (right panel), are depicted
as functions of the time  using the initial GM profile
 given by Eq. (\ref{gaus}) with $X=0.1$. }
\label{figu2}
\end{figure}

For larger  $X$ values we do not observe such pure solitonic fronts, since 
the phase accumulation distributed  along each notch now spreads around a vortex state. 
  Such combined soliton-vortex quasiparticles  are usually  called solitonic vortices  or 
 svortices \cite{brand02}. 
We have found that these vortices move with the soliton angular  
velocity  along a circular trajectory of a radius slightly larger than $r_0 $, 
except near the zone where the collision takes place, where they perform a complicated dynamics. 
 In Fig.~\ref{figu2} we show the results of a GP simulation using an initial GM order parameter
 given by Eq. (\ref{gaus}) 
with $X=0.1$.   The angular position of the 
soliton located at $x>0$ (left panel) and the phase difference $\Delta \phi $ 
(right panel), are depicted as functions of time.
 We observe an increasing
absolute value of the slope of $\theta(t)$ after each collision. Also 
the maximum absolute value $ \theta_M $ of the angle  
$ \theta$ keeps growing along the evolution.  
For example,  for the first and third collision we obtained $ \theta_M/ \pi= 0.44485$  and 
$ \theta_M/ \pi = 0.4514$, respectively.
  After many oscillations, at $ t\simeq 470 $ ms
when  $ \theta_M/ \pi = 0.5$ is attained, it may be seen that the system enters a different regime
that will be described in the next subsection.
Such a transition occurs when
the soliton exceeds  a critical velocity  $ v_c \simeq 0.8\, \mu \text{m}/ \text{ms}$.

\begin{figure}
\includegraphics{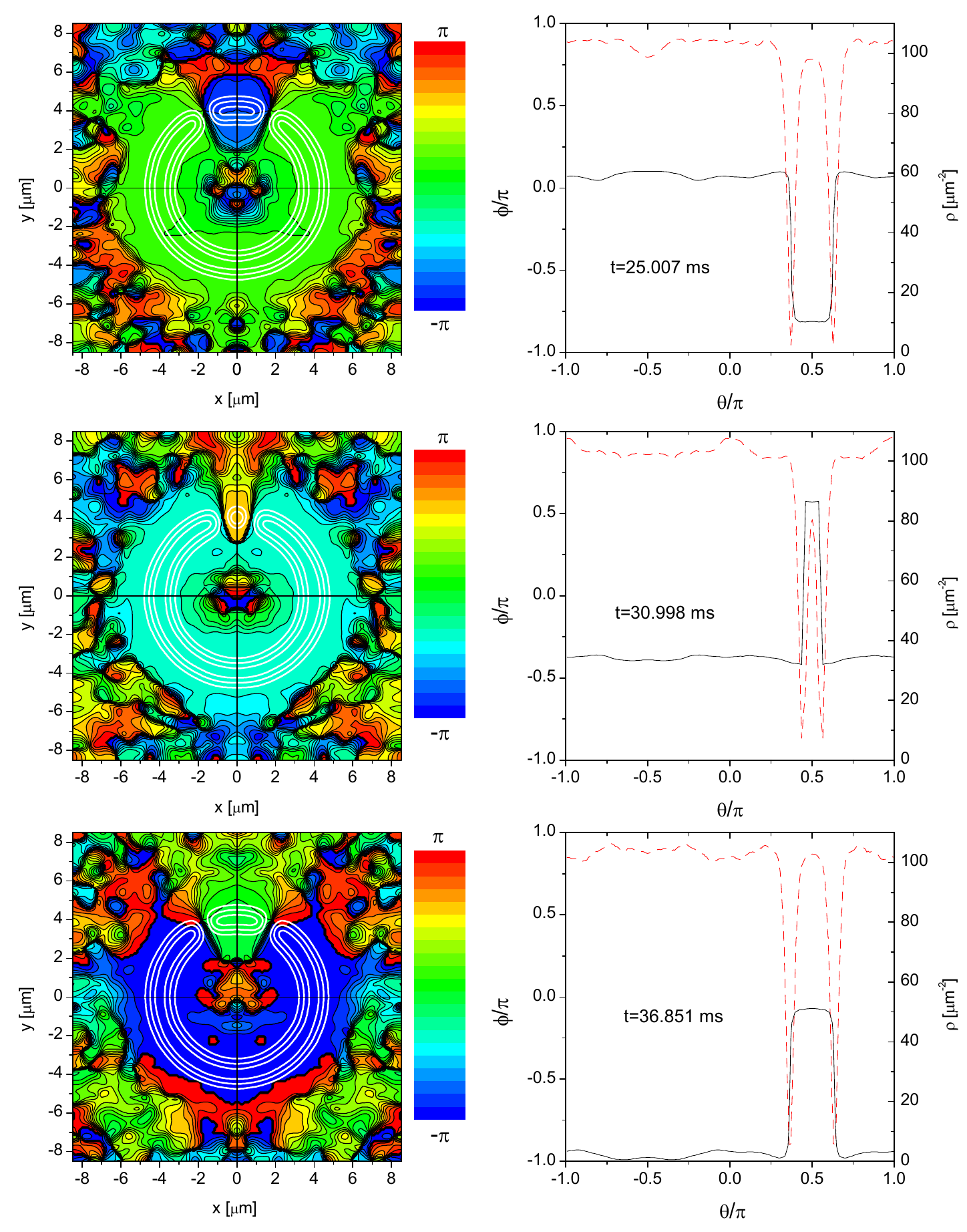}
\caption{(Color online) 
Snapshots before (top panels), during (middle panels), and after (bottom panels)  the first soliton
collision for the same initial state of 
Fig.~\ref{figu2}. In the left panels we show the phase distribution (colors) and density isocontours 
(white solid lines), while in the right panels we depict the
corresponding angular distribution of particle density  (red dashed line)  and phase 
(black solid line) at $r= 4 \, \mu$m.}
\label{figu3}
\end{figure}

In Fig.~\ref{figu3} we show snapshots of the phase and particle density around the first collision 
for different times. 
It may be seen that in the collision time  (middle  panels), the density presents 
two well separated notches forming an angle $  \Delta \theta \simeq 0.11 \pi$.  
In this case, the  passage of vortices along the notches, in the radial direction,
 is responsible for the velocity field inversion, as also has been observed in 
the self-trapping regime 
for a double-well system, where the vortices move along
the junction \cite{stvor,*stvor1}. In particular, the velocity field inversion in this collision may be produced
by either the motion along the notch of
an external  ($ r > r_0 $) counterclockwise
vortex  in the negative radial direction, which is the present case,  or by an inner ($ r < r_0 $) clockwise vortex,
passing through the notch towards the outside region. 
We note that such  a dynamics  has no physical analogue in pure 1D systems as they are
vortex-free.
In the  top-left panel of Fig.~\ref{figu3},
we may see that before the collision
 the vortex associated with the soliton at the right has a counterclockwise 
vorticity and, as it is located at
$ r >   r_0 $, 
the net flux of particles across the corresponding density notch points  clockwise, as expected. 
On the other hand, after the collision (bottom-left panel), the
new arising vortex presents a clockwise vorticity, consistent with the change of direction of the
velocity field.
A similar behavior  has been observed for the different initial conditions
up to
soliton velocities around $ v_c$.

Finally we want to note that in the presence of vortices, the  density does not necessarily vanish  during the collision  at $r=r_0$, 
 as seen in the middle-right panel
 of Fig.~\ref{figu3}. One can only assure that  the density  vanishes at the points where the vortices are located.
As can also be observed from the right panels of Fig.~\ref{figu3}, one has
  $|\Delta\phi| \simeq \pi $  at the collision (middle panel), whereas before and after the 
collision smaller values of $|\Delta\phi|  $ are obtained. 
In addition, the change of  sign that $\Delta\phi $ undergoes  between the upper 
an lower panels is consistent with 
the inversion of the soliton velocity.

\subsection{ Soliton transmission at the collision ($ 0.5 < X < 1 $)}

As we  have pointed out in the previous subsection,  for velocities larger than  
$  v_c  \simeq  0.8\, \mu  \text{m}/ \text{ms}$ the system enters a different
regime. And in fact, for $ X> 0.5 $, we have found that the initial soliton velocity  exceeds   $v_c$, 
and thus the whole evolution lies within this regime.
Similarly to what happens in infinite 1D systems for    $ v_s/c  >  0.5$   \cite{theo10}, 
we shall see that in such a regime 
the solitons can be thought of as being transmitted through each other, as they completely
overlap during the collision.
We recall that in
infinite 1D systems a single notch of vanishing density is produced at the collisions with 
 $ v_s/c  = 0.5$  \cite{rev10,theo10},
which is characteristic of a 
double root of a real  order parameter preserving the same sign on the overall space.
Whereas for $ v_s/c  > 0.5$, such a single notch has a nonvanishing density and the colliding solitons
have been interpreted as transmitting 
through each other \cite{theo10}.
It is worthwhile mentioning that in experimental works on bright solitons \cite{nguyen14},
it has been shown that the solitons
pass through one another and emerge from the collision unaltered in shape, amplitude, or 
velocity, but with a new trajectory. 
Similar qualitative features have been observed in our case, but this time
accompanied with an associated vortex dynamics. 
However, it is important to note that 
we have found that the soliton velocity slightly increases during the evolution, becoming
 after many collisions appreciably larger than the initial value. 

\begin{figure}
\includegraphics{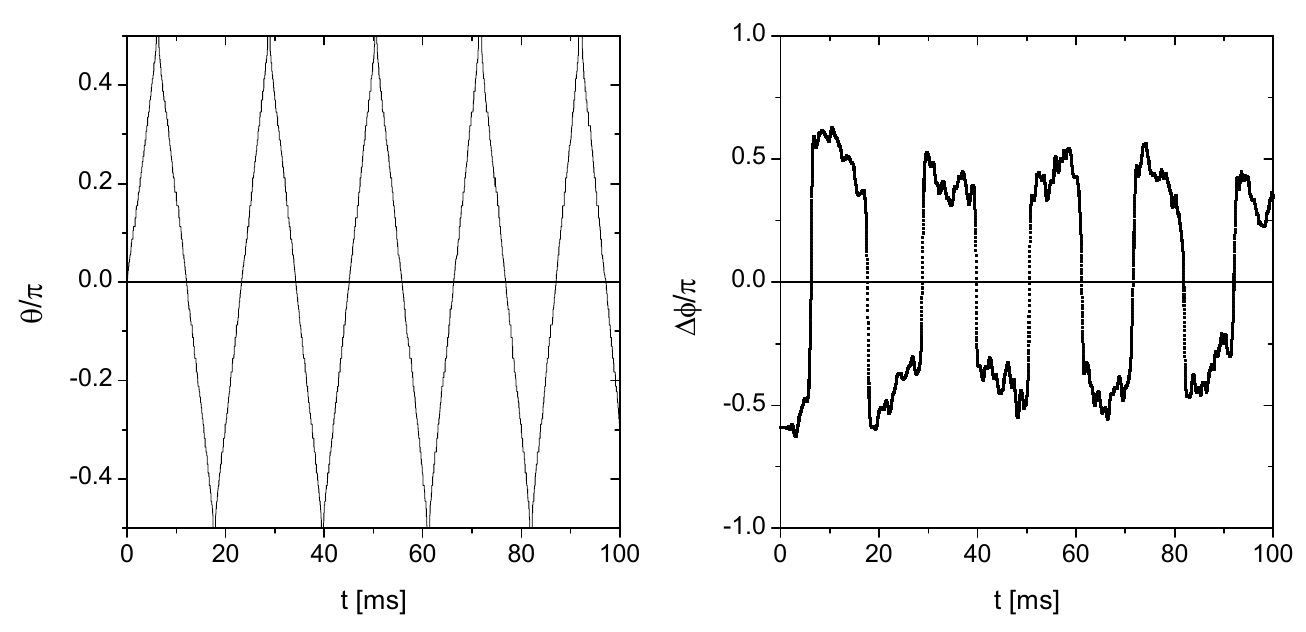}
\caption{
Angular position (left panel) of the soliton located at $x > 0$
and the phase difference $\Delta \phi $ (right panel), are depicted
as functions of the time for an initial GM profile
 given by Eq. (\ref{gaus}) with $X=0.6$.
  }
\label{figu4}
\end{figure}

 In the left panel of Fig.~\ref{figu4},  we show the angular position 
 of the soliton located at $x>0$
 as a function of time for $ X=0.6 $, where it may be seen that at each collision
 the maximum absolute value reaches the value
$ \pi/2 $, which means that both solitons approach each other 
until they completely overlap just at the collision time.

On the other hand, the time evolution of
the phase difference is shown at the right panel of Fig.~\ref{figu4}, 
where we may observe that in contrast to the  regime with $v_s< v_c$,
the phase difference remains bounded. 
Particularly, one can estimate such an  upper bound  from
the formula $ v_s/ c =\cos(\Delta\phi/2)$ derived for an infinite 1D system,  assuming that the transition occurs for 
 $v_s = 0.5  \, c $ 
which yields $\Delta\phi/\pi=2/3$,
a value that agrees very well with that observed at the  right panel of   Fig.~\ref{figu2}. 
Moreover, just at the collision time, the phase difference
also shows a quite distinct behavior for solitons moving slower or faster than
the critical velocity. That is, we have observed that $|\Delta\phi|$ reaches the value $\pi$ at collisions
with $v_s< v_c$, whereas it goes to zero for soliton
velocities above such a value. This behavior is clearly shown on the right panels of Figs. \ref{figu2}  
and  \ref{figu4}.

\begin{figure}
\includegraphics{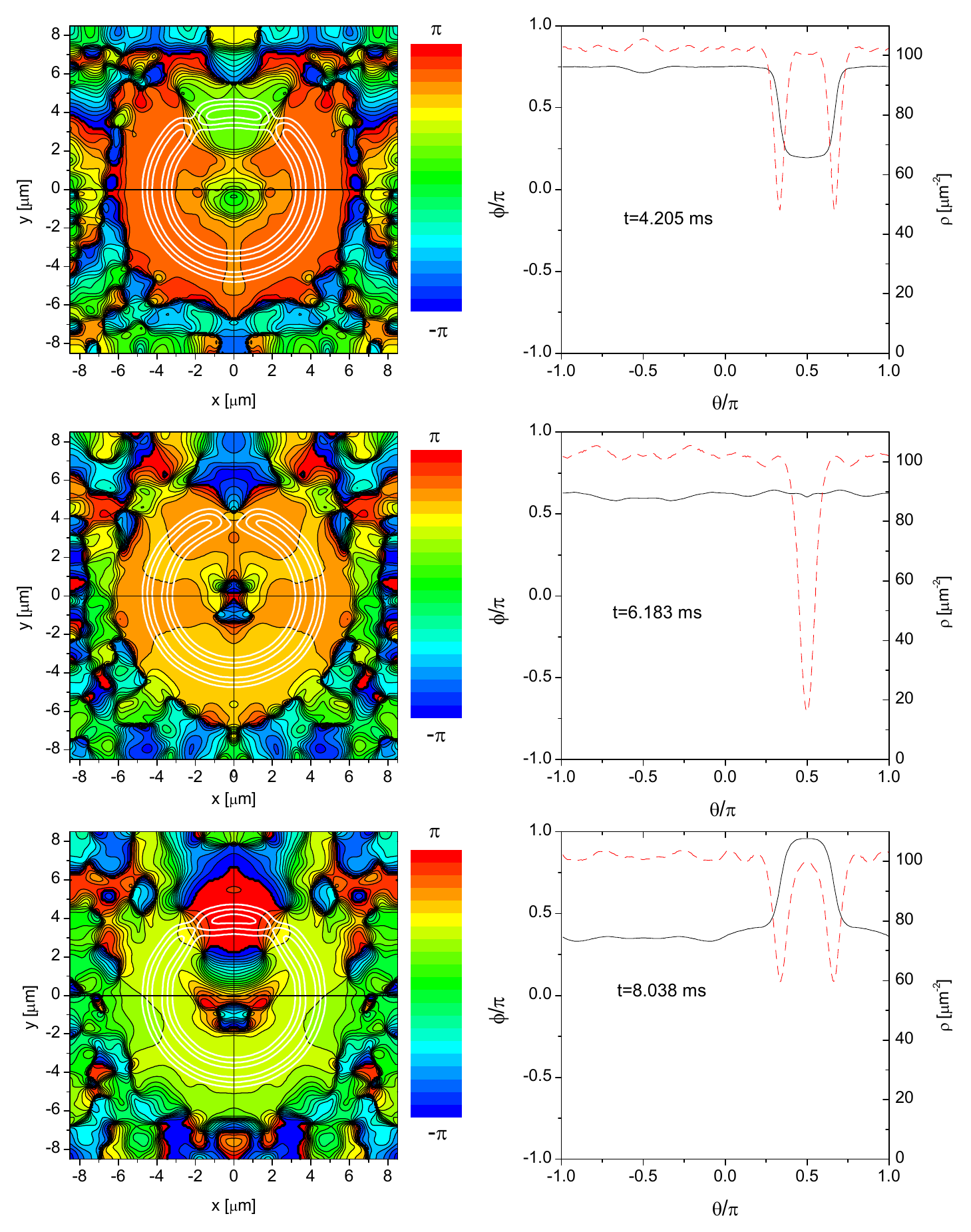}
\caption{(Color online) Snapshots before (top panels), during (middle panels), and after (bottom panels)  the first soliton
collision for the same initial state of 
Fig.~\ref{figu4}. In the left panels we show the phase distribution (colors) and density isocontours 
(white solid lines), while in the right panels we depict the
corresponding angular distribution of particle density  (red dashed line)  and phase 
(black solid line) at $r= 4 \, \mu$m.}
\label{figu6}
\end{figure}

The dynamics in this regime is again ruled by vortices, 
which precede around the condensate, except near the collision, where 
once more a complex vortex dynamics takes place. 
It can be seen in the top-left panel of Fig.~\ref{figu6},
that the vortices are located farther from the center  than in the
previous case, which is in accordance with a larger vortex velocity \cite{velo,*velo1},
  as the soliton also moves faster.

 During the collision we have observed
the annihilation of the vortex and the antivortex  coming with each soliton. 
The counterclockwise vortex at $x>0$, viewed at the top-left panel,
annihilates with the clockwise vortex located at $x<0$. In fact, in the middle-left panel it may be seen that
both vortices  have completely disappeared. 
Just after the collision, a new vortex-antivortex pair is generated at each notch. The vortex and  antivortex  forming the pair
 rapidly
separate from each other, as can  be seen in the bottom-left panel of Fig.~\ref{figu6}, 
and   from analyzing subsequent times,  we have observed that the inner vortex becomes absorbed 
towards the central
region.  On the other hand, the outer vortices with vorticities
opposite to those before the collision, keep ruling the soliton dynamics.

By comparing the right panels of Fig.~\ref{figu6},
  it can be seen that the density   reaches
the minimum, yet nonvanishing, value  at the collision time (middle panel) when the notches completely overlap.
On the other hand,    $\Delta\phi   $ changes its sign between the upper and lower panels, while
 the   phase is almost uniform in the middle one,  yielding  $ \Delta\phi = 0 $.

 Finally we want to note that our numerical results are consistent with the fact that faster solitons have smaller depths,
as can be seen from the density plots of the right panels of Figs. \ref{figu3} and  \ref{figu6}.

\section{Dependence of the physical quantities on $X$}
In this section we will study the dependence on $X$ of the initial soliton energy and velocity, and also
the position of  turning points at the collisions, comparing all these results to the infinite 1D case.
It is worthwhile mentioning that we will disregard the presence of vortices in the analytical calculations.

\subsection{Energy}

We first recall that
in an infinite 1D system, the single dark soliton order parameter has an analytical expression which
is characterized by the parameter $ X = v_s/c$ \cite{tsu71,libro,rev10}. 
 In addition, it is easy to verify that the associated energy can be written as
 $ E_{1D}= (4/3) \hbar n_0 c (1-X^2)^{3/2}$ \cite{konotop04,libro,rev10},
where $n_0$ denotes the background density.

In this subsection we will study the dependence on $X$ of the double-notch soliton energy for 
the PBS, GM and PG initial states. In particular, for the GM order parameter (given by Eq. (\ref{gaus})), we may also obtain
an expression for the energy  by performing the same type of  approximations in the integrals as 
we have done to derive Eq. (\ref{densidad}).
Taking into account such considerations, we may  finally obtain the soliton energy by  evaluating the kinetic, trap and interaction energy 
terms, and then subtracting
the  ground state energy ($ E_0$), 
\begin{equation}
E_S(X) = E(X)- E_0  \simeq  \frac{4(1+\sqrt{2})}{3 \sqrt{2}} \frac{g \,
 n(X)^2 B}{k r_0}  (1- X^2)^{3/2}-\frac{4}{ \pi  \sqrt{2}} \frac{g \, n(X)^2 B}{(k r_0)^2}  (1- X^2)
\label{energyv}
\end{equation}
where $B= r_0^2 \sqrt{ \pi/ \gamma}$. We observe a similar dependence on $X$ to
 the infinite 1D case, but it is important
to stress that in this case we do not have such a simple relation between $X$ and the soliton velocity. 

\begin{figure}
\includegraphics{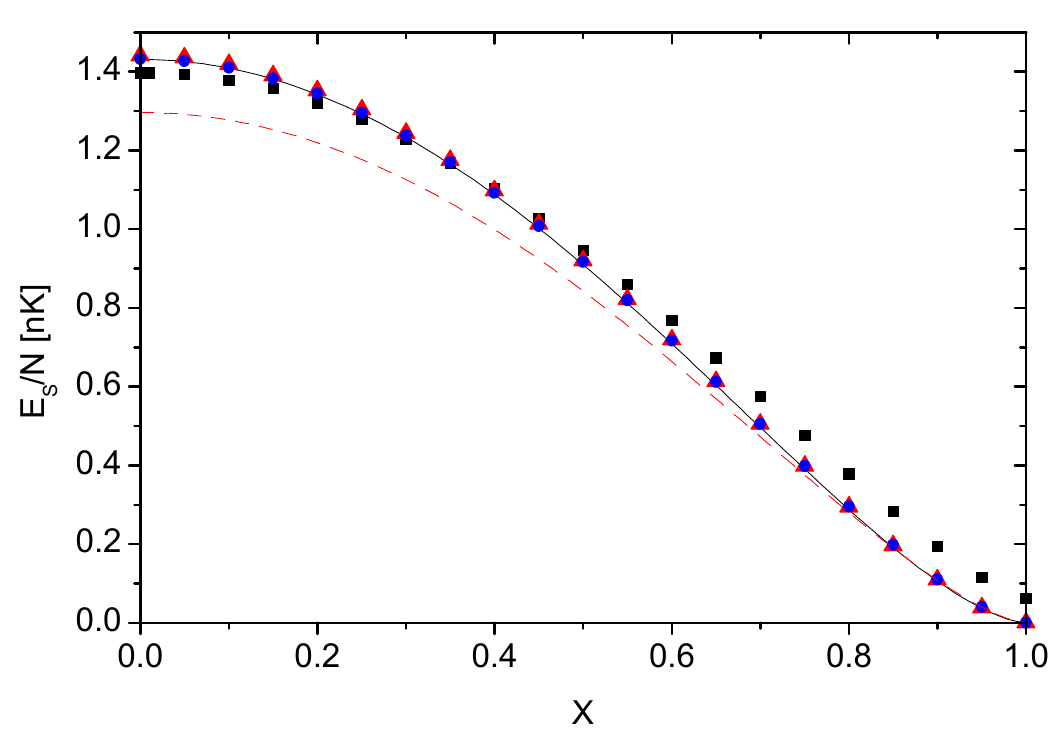}
\caption{(Color online) Soliton  energy per particle as a function of $X$. The solid black line corresponds to 
Eq.~(\ref{energyv}); the 
 black squares, blue circles, and red triangles correspond to numerical calculations using  PBS, GM,  and PG states, respectively. 
The red dashed line corresponds to an infinite 1D system.}
\label{energy}
\end{figure}

 In Fig.~\ref{energy}  we display our numerical calculations of the energy per particle, together with
the estimate arising from  Eq. (\ref{energyv}).
Note that  for $X$ approaching zero (unity), the energy 
should be better described by the PBS (PG) order parameter. In this context, we observe that
the energy arising from
the GM order parameter (\ref{gaus}) and the corresponding estimate (\ref{energyv})
accurately reproduce the PG results. 
We also show in Fig.~\ref{energy} 
the $X$ dependence of the 1D soliton energy $ E_{1D} $, which turns out to be qualitatively
similar to that arising from Eq. (\ref{energyv}).

\subsection{Soliton velocity}

\begin{figure}
\includegraphics{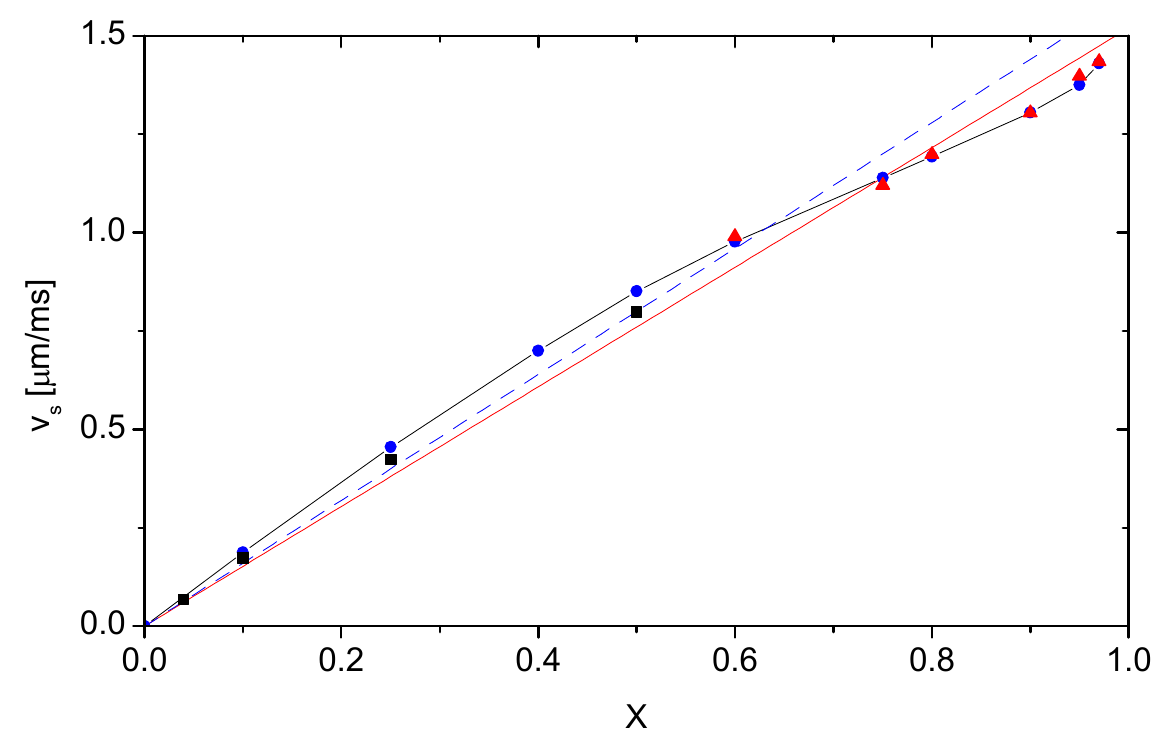}
\caption{(Color online) The soliton initial 
velocity for the  PBS (black squares), GM  (blue circles), and PG (red triangles) states depicted  as a 
 function of $X$. The black solid line has been drawn to guide the eye on the GM results.
The linear functions $  c_i X $ are depicted as a red solid line ($i=0$) and as a dashed blue line ($i=1$),
 where $c_i$ is an estimate of the 
sound velocity at either  the ground ($i=0$) or the black soliton ($i=1$) states. 
}
\label{figu7}
\end{figure}

Here we will analyze the relation between the parameter $X$ 
and the initial  soliton velocity.
To elucidate whether we can assign to the parameter $X$
the same physical meaning as in the infinite 1D case, namely the ratio between soliton and
sound speeds, we depict  in Fig.~\ref{figu7} the soliton 
velocity before the first collision as a  function of $X$. 
We recall that as seen in  the previous section,  the soliton speed may  increase along the time evolution.
However, it is worthwhile noticing that the soliton
velocities before the first collision turn out to be quite similar for
the three different kinds of initial order parameters.
 We have also plotted in Fig.~\ref{figu7} the linear functions $ f_i(X)=c_i X $, 
where $c_i$ is an estimate of the 
sound velocity obtained by tracking the motion of 
a small localized density perturbation, in either the ground  or the  black soliton 
   states, which 
yielded $c_0= 1.52\, \mu \text{m} / \text{ms} $ and $c_1=1.6\, \mu \text{m}/ \text{ms}$, respectively.
Such functions $ f_i(X) $ represent estimates of the soliton velocity assuming a linear dependence
on $X$ as predicted for an infinite 1D system, using both extremes   $ c_i $  of the  sound velocity.
We want to remark that as  the size of the condensate is 
finite, the background density changes with the depth of the notch,
and thus the sound velocity also varies. 
In particular,  the sound velocity of the ground (black soliton) state
turns out to be the smallest (largest) one.
Therefore, for any intermediate $X$ value, the sound velocity $c(X)$ should verify 
$ c_0 \leq c(X) \leq  c_1 $.

From  the  results depicted in  Fig.~\ref{figu7}, we may certainly conclude that the parameter $X$
can roughly be
approximated by the ratio of the soliton velocity  and a mean sound speed between  $c_0$ and  $c_1$.
A possible source of discrepancy in this respect could arise from
the fact that even before the first collision, the soliton velocity undergoes a gradual growth due 
to energy dissipation.  In addition, for large soliton velocities the presence of vortices may 
also be affecting the results.

It is worth to notice that, as can be seen in Fig.~\ref {figu7}, within the dispersion of data given by the 
different approaches,  the soliton velocity 
 lies  around $ v \simeq 0.8 \mu m / ms$ for $ X=0.5$,  which
is consistent with  the classification of the dynamics  we have performed  using either  $X < 0.5 $ or $X > 0.5 $.

\subsection{Turning points}
In a linear homogeneous 1D system, the
turning points for low speed solitons ($ X=v_s/c \leq 1/2$) have been calculated analytically \cite{theo10}, yielding 
the following expression for the minimum distance $d$ reached in a symmetric collision:
\begin{equation}
d=\frac{ \zeta }{  \sqrt{1- X^2}}   \text{ cosh}^{-1}( \frac{1}{X} - 2 X)  ,
\label{acercamientoz0}
\end{equation}
where $\zeta$ denotes the healing length. 

Assuming one may apply formula (\ref{acercamientoz0})
to  our toroidal configuration to estimate the turning angle $\theta_M$ for the first collision of the soliton located
at $x>0$, one obtains
\begin{equation}
\theta_M = \frac{\pi}{2} - \frac{ 1 }{  2  k r_0 \, \sqrt{1- X^2}}   \text{ cosh}^{-1}( \frac{1}{X} - 2 X)  ,
\label{turningpoint}
\end{equation}
\begin{figure}
\includegraphics{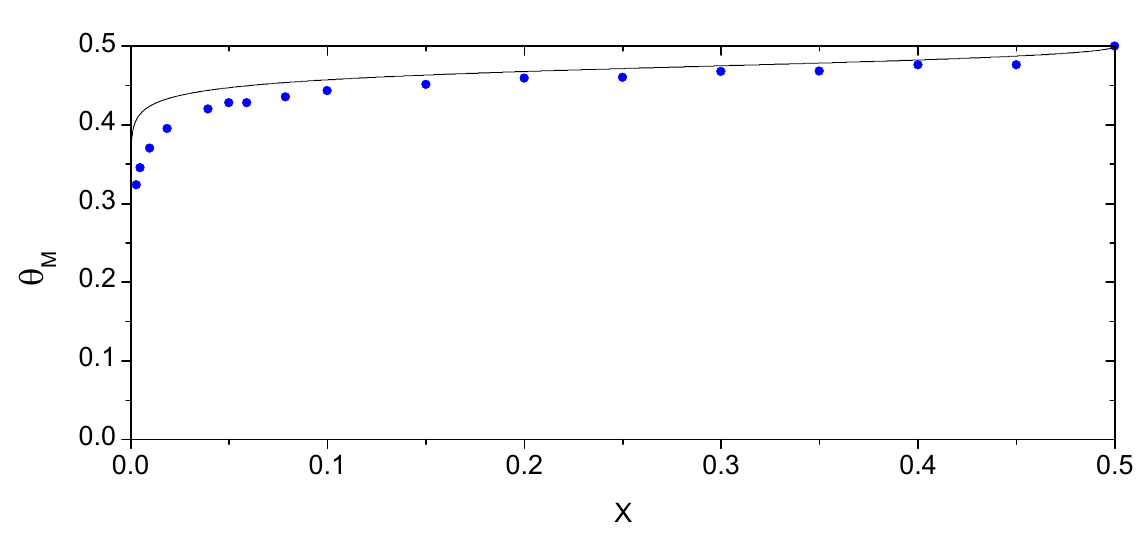}
\caption{The turning points arising from GP simulations using the GM order parameter (blue circles)
and the estimate given by Eq. (\ref{turningpoint}) (black solid line)
 are depicted as a function of $X$.}
\label{figu11}
\end{figure}
whose results are depicted in
Fig.~\ref{figu11}. In the same figure we show the corresponding GP simulation results
using the   GM order parameter as the initial state. We do not depict  the points arising from PBS and PG states
because the three approaches threw practically the same results.
By comparing the solid line with the distribution of points, a  
qualitative similar behavior is observed, with a discrepancy that may be adjudicated  to the 
difference in the  geometry between the  torus and an infinite  1D system. 

Finally, we note that 
a repulsive potential between solitons, from which an estimate for the turning points
can be derived, has been proposed for the 1D system in the limit $ X\ll 1/2$ \cite{theo10}. 
Assuming that such a potential also rules the dynamics of our system, 
we derived another estimate of the turning angle, which yielded practically the same
results of Eq. (\ref{turningpoint}) for  $ X \lesssim 0.2 $.

\section{Conclusions}
We have observed quite
different characteristics of the collision processes depending on the soliton velocities.
For very slow speeds, the solitons remain  as long living defects whose dynamics presents almost 
constant velocities, except within a narrow 
region where the collision takes place. In this process, the colliding solitons lower their velocity to zero 
and continue moving in the opposite direction 
with the same speed that had before the collision. Therefore, such an event can be considered as
an elastic collision, and hence, in energy terms, the soliton  dynamics seems to be non dissipative. 
On the other hand, the turning points are always located at the same angle during the time 
evolution, with
well separated notches and fronts sharp and straight. There is no evidence of vortex 
generation near the condensate that can affect the soliton dynamics. 

 For faster solitons,  the dynamics is ruled by 
vortices which define svortices structures. Such vortices  precede around 
the condensate, except within a small region where the collision takes place,
where they perform a more complex dynamics around the notches. 
In these processes the solitons lose energy, as the outgoing
velocity turns out to be larger than the incoming one. 

For soliton  speeds smaller than approximately  half the sound velocity, the density dips remain
separated during the collision and the passage of vortices along each notch, in the radial direction, causes the
 inversion of the local velocity field.
 We find that the angular distance of closest proximity between the colliding solitons
 decreases with the increasing soliton velocity. 
   Taking into account  that the notches are well separated during the collision, 
in accordance with the infinite 1D case \cite{theo10} we may say that
 the solitons  are reflected by each other.
However it is important  to recall that the outgoing vortex has the opposite  circulation to the incoming one.

A different regime arises for soliton speeds larger than around  half  the sound velocity.
Namely, as in the 1D case \cite{theo10}, the density dips become completely overlapped at the collision
and the  solitons are transmitted through each other. In this case
the vortex dynamics during the collision turns out to be simpler than in the preceding case, since
the vortices that come from each soliton front annihilate 
with each other, and afterwards new vortices are generated that start ruling the subsequent dynamics. 

As a general remark we want to notice that, as we have seen in most cases, during the evolution
the soliton increases its velocity and thus 
reduces its energy. It is then natural to wonder where such
a released energy is being transferred, given that it is well known that the
time-dependent GP equation conserves the energy of the entire system. 
From the analysis of the whole dynamics, we have observed that 
such a soliton energy  dissipation is accompanied by  an  increasing number of vortices 
that become gathered together in the inner region of the torus, 
 along with the appearance of  a large amount of density fluctuations. These excitations  may  be viewed, at an early stage, 
  in the left panels of 
 Figs.~\ref{figu3} and   \ref{figu6}.
Hence, we may conclude that part of  the initial soliton energy is being 
transferred to the increasing number of  vortices and density fluctuations.

From 
the study of the energy, soliton velocity and turning points before the onset of  dissipation,
we have found that the behaviors of  these quantities as  functions of $X$
 qualitatively resemble those of  the infinite 1D case for $X=v_s/c$.

Finally, we wish to point out that a corresponding
experimental set up of our system seems to be definitely within the reach
of the present investigations. Particularly,
the toroidal condensate we have proposed has already been experimentally achieved  \cite{boshier13},
while our solitonic initial profiles could eventually be generated by 
standard phase and/or density  manipulation methods \cite{becker08}.

{\it Note added.} After the initial submission of this paper a preprint appeared \cite{gallu}
where the authors show the feasibility of experimental
generation of counter-propagating solitons, moving at velocities
above half the sound speed in a similar toroidal geometry.

\acknowledgments
 PC acknowledges financial support from ANCyPT through Grant No.  PICT 2011-01217.
HMC acknowledges CONICET for financial support under Grant No. 
PIP 11420100100083.

\providecommand{\noopsort}[1]{}\providecommand{\singleletter}[1]{#1}%

\end{document}